# Human Behavior Plasticity Measured in Speech Epochs: A Proof-Of-Concept Study

Vikram Jakkamsetti [1], Clinton Broyles [2], Frank Buttafarro [2], Clint R. Myers [2], Sophia Kwong Myers [2]

[1] Department of Neurology, The University of Texas Southwestern Medical Center, Dallas, Texas, USA; [2] Independent Researcher, Dallas, Texas, USA

**Correspondence**:
Vikram Jakkamsetti, MD, PhD
Instructor, Department of Neurology
UT Southwestern Medical Center
5323 Harry Hines Blvd. Mail code 8813, Dallas, TX 75390-8813
E-mail: Vikram.Jakkamsetti@UTSouthwestern.edu

**ABSTRACT**
Human behavior training in improvisational theater has shown extensive behavioral and health benefits. Improved empathy measures in medical students, improved behavior outcome in patients with autism and a reduced recidivism rate are among the many benefits attributed to improvisational theater training. However, measuring tangible outcomes of changed behavior is challenging and usually requires multiple sessions or months of training. One of the principal tenets of improvisational theater is to actively listen and collaboratively allow a scene partner to talk and provide equal input as needed. Here we measured human speech epochs and asked if a month of weekly improvisational theater training would reflect in speech epoch assays. We found a significant decrease in speech epoch durations after one month of weekly training. There was no change in epoch durations on the same day, suggesting it to be a stable parameter over a day, but amenable to modulation over a month of training. The overall rhythm of speech epochs remained unchanged but durations of discrete low frequency speech volleys increased. Moreover, training improved regularity of adjacent speech epoch durations, suggesting a better sharing of speaking focus by matching speech time to the previous speech epoch duration. Our assay provides a proof-of-concept study of empathy-relevant tractable speech epoch parameters associated with changes in a month after human behavior training.

**INTRODUCTION**

Adaptation to the surrounding environment for survival with behavioral plasticity is conserved across species [1-3]. Behavioral plasticity to simpler paradigms in less evolved species is easier to measure. For example fear conditioning in Aplysia (sea slugs) is amenable for measurement, yielding multiple reductive insights [4]. Some higher order behaviors and related plasticity in rodents can be measured using tracking tools and machine-learning analysis [5]. However more complex behaviors in higher evolved species like humans are challenging to measure in a manner conducive to reflecting modulation or training associated effects.

Human behavior training in improvisational theater has shown clear benefits in outcome behaviors measured. Medical students trained in improvisation had higher empathy associated behavior with improved communication with patients [6,7]. Improvisational theater has shown promise in behavioral therapy for patients with autism [8]. One striking example is of marked reduction in recidivism (from ~50 % to ~11%) for those involved in a years-long Actors' Gang's Prison Project supported by the actor Tim Robbins [9]. Assaying outcomes like recidivism is effective but take considerable time and numbers to be statistically meaningful. Having behavior-relevant measures that are more immediate and objective would be helpful.

Improvisatonial theater varies from conventional drama theater in forgoing a script to respond in the moment and being unpredictable. However, one core training principle involves a "yes and" to one's scene partner wherein one accepts an offered dialogue/move however improbable ("Son, we are on the moon") and adds more context to it ("Yes, and bring-your-son-to-work-day is so much fun"). The training encourages active listening and collaboration and the allowing of each participant to be given the opportunity to be heard. Training often progresses from participants talking over each other in initial sessions, to learn to listen and intuitively allow equal participation if the scene calls for it. We theorized that measuring participant's speech epochs could track an evolution in improvisational behavior. Specifically, we hypothesized that speech epoch durations would be longer at the onset of training due to the occurrence of multiple voices talking over each other and shorten with a month of weekly training. Moreover, we theorized that training would improve the rhythm of speech epoch occurences. We found proof-of-concept results that indicate a decrease in speech epoch durations and an increase in their regularity with Improvisation theater training.

**METHODS**

*Improvisational theater*: Long-form improvisational theater was practiced among five members of the troupe Hippo Campus starting at 11 am each Saturday for a month at the same location. Each session lasted 17-22 minutes and one to two "jam" sessions were planned. Each session began with a random word picked from the New York Times as an inspiration for the start of a scene and subsequent edits and scenes were inspired from information generated in the first scene.

*Data acquisition and analysis*: Audio recordings were conducted on an iphone 7 and the .mp4 files were accessed by scripts written in MATLAB as an amplitude time-series for further analysis. The sampling frequency was 44.1 kHz. Analysis was similar to that done in a recent study for an amplitude-based detection of a time-series signal [10]. In brief, an envelope contour of the audio signal was generated and any signal that was greater than 3 times the baseline value marked the occurrence of a speech epoch. Communication between scene partners can in theory involve sound that is not speech, but for the purposes of this manuscript all sound initiations are labeled as speech sounds. Power spectral density was calculated for each 20 second segment of an audio recording with MATLAB's available fast fourier transform function. Speech volleys at different frequencies were captured for the upper amplitude of the speech amplitude time-series as done earlier by us for electrophysiological signals[10]. A speech volley of equal to or greater than one second and occuring more than one second after the last speech volley was flagged for analysis. CV2, an extension of coefficient of variation and measure of irregularity of adjacent time epochs, was calculated as done earlier for determining brain neuron firing regularity[11]. CV2 is determined by calculating the the absolute difference between each adjacent pair of epoch durations and dividing that by their mean. The formula for CV2 is given below:

$$\text{CV2}_i = \frac{|\text{Epoch}_{i+1} - \text{Epoch}_i|}{(\text{Epoch}_{i+1} + \text{Epoch}_i)/2}$$

where "Epoch" represents each $i^{th}$ speech epoch duration.
*Statistical analysis*: Prism 6.0 (Graphpad software) was employed to generate frequency distributions and conduct an ANOVA with posthoc Tukey's test for multiple comparisons.

**RESULTS**

*Extraction of speech epochs*
   Visual inspection of amplitude time-series plots of audio recordings showed distinct peaks associated with human speech occuring prominently above the baseline (Figure 1A). All algorithm detected speech epochs were confirmed as associated with human speech sounds with simultaneous playback of the audio recording for each given twenty second plot.

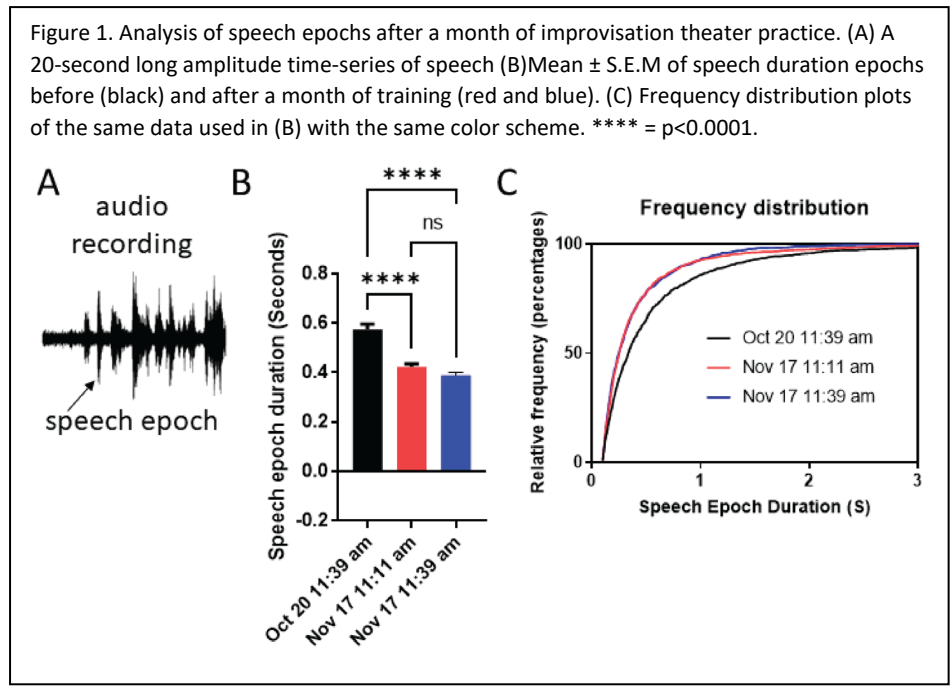

Figure 1. Analysis of speech epochs after a month of improvisation theater practice. (A) A 20-second long amplitude time-series of speech (B)Mean ± S.E.M of speech duration epochs before (black) and after a month of training (red and blue). (C) Frequency distribution plots of the same data used in (B) with the same color scheme. **** = p<0.0001.

*Behavioral plasticity in speech epoch duration with one month improvisational theater practice*
   One core tenet of long form improv is to avoid negating scene partners and to "yes and" and allow participation of members in the common interest of the scene. We hypothesized that with practice as a group, there would be an intuitive give-and-take with less talking over and interruptions. Thus we hypothesized that with fewer examples of participants talking over each other, speech epoch durations would get shorter. We found a significant decrease in speech epoch durations after one month of weekly training (Figure 1B-C). There was no difference in speech epoch durations for sessions collected on the same day. Our findings indicate that speech epoch durations were stable across a day but modulable with training over a month.

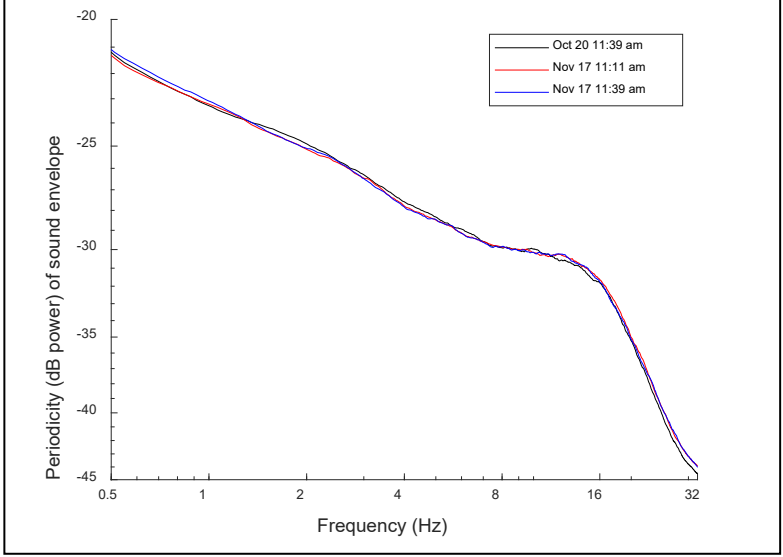

Figure 2. Average relative power spectral density of the envelope of the amplitude time-series of each Improv session

*Power spectral density as a measure of speech epoch rhythm does not change with training*

We predicted that with trainig and familiarity of troupe member proclivities, the exchange of speech epochs between players will have less overlap and there will arise a rhythm of give-and-taking focus between players in a scene. Power spectral density analysis is often used to examine for repeating rhythmic patterns at different frequencies in time series data. Since the ampiltude of a time series signal affects power spectral density, we chose to use relative power spectral density to adjust for the possibility of changes in loudness of the recordings over sessions. We asked if the average relative power spectral density of the amplitude envelope of speech showed any differences with training. Contrary to our expectation, there was no change in power spectral density (Figure 2), suggesting no change in overall rhythm of speech epoch occurences.

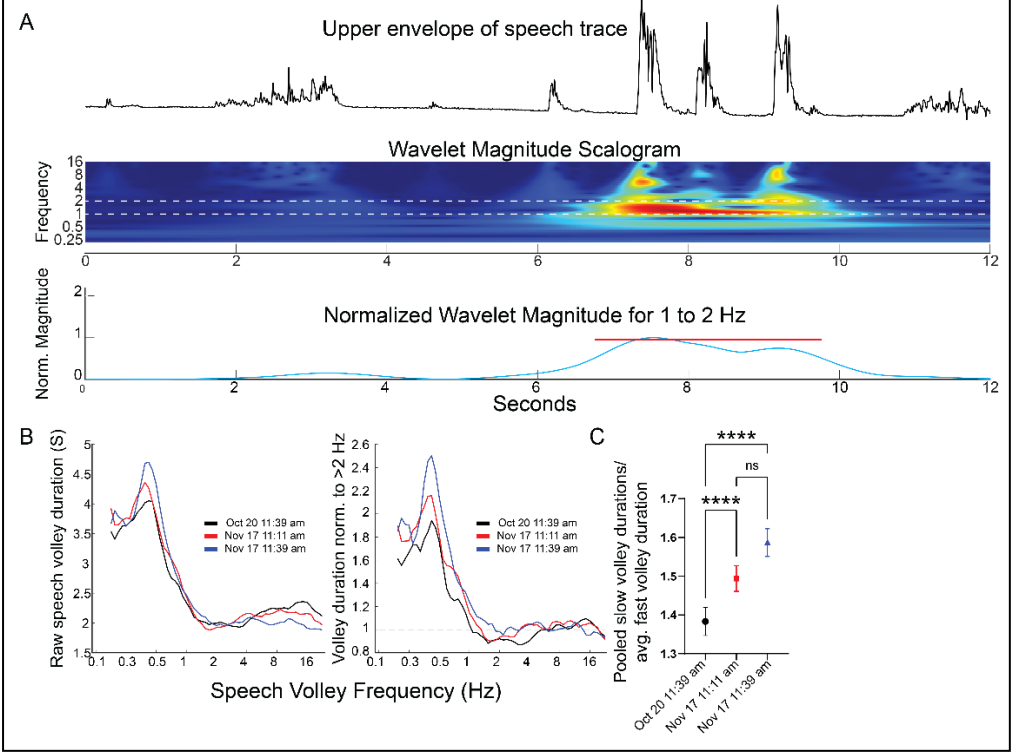

Figure 3. Speech volley capture and duration. (A) Upper envelope of speech amplitude time-series recording (*upper*), its visualization using a wavelet transform (*middle*) and automated capture of a 1 to 2 Hz speech volley (*lower*) (B)Average raw and normalized speech volley duration for different frequencies (C) pooled normalized volley durations. **** = p<0.0001.

*Slow Speech Volley durations increase with training*
We reasoned that power spectral density can only measure a rhythm that occurs for the majority of the recorded signal. Similar to a recent study published by us wherein infrequent oscillations were missed by power spectral density analysis, we postulated that speech volley occurences would be intermittent and missed by power spectral density analysis. Indeed, an analysis capturing discrete speech volleys (Figure 3A) indicated the predominance of slow speech volleys with speech epochs occuring once every 2 to 4 seconds (Figure 3B) for all recordings. We presumed that high-frequency volleys (>2 Hz) would reflect less deliberate and more chaotic speech interactions. Normalizing the data to the average high-frequency speech volley provided a parameter that showed significant improvement in slow volley durations with training (Figure 3C).

*Training improves regularity of adjacent speech epoch durations*

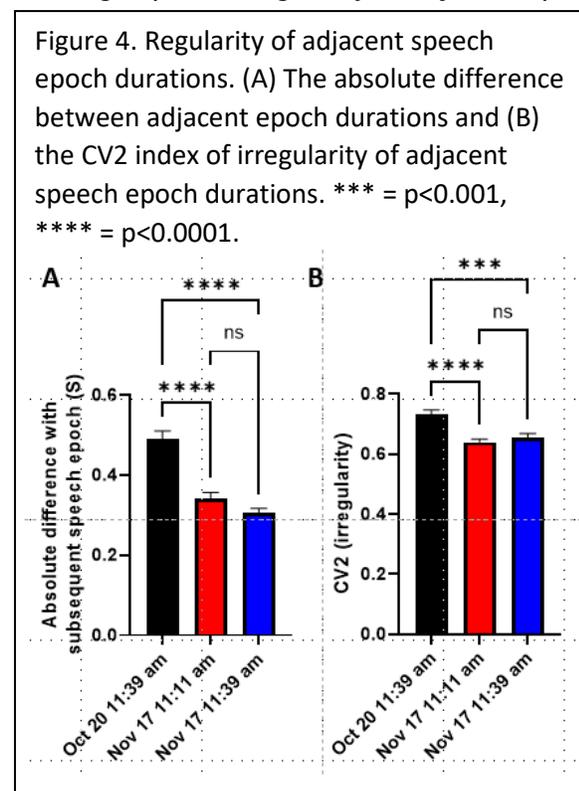

Figure 4. Regularity of adjacent speech epoch durations. (A) The absolute difference between adjacent epoch durations and (B) the CV2 index of irregularity of adjacent speech epoch durations. *** = p<0.001, **** = p<0.0001.

Improv practice in general involves sharing focus on stage without any particular player dominating speaking time. We asked if training to give equal focus to troupe members would reflect in matching speech epoch durations between players in a scene. We found that the average absolute time difference between adjacent speech epoch durations was signifcantly lower after training (Figure 4A). A shorter speech epoch as seen in Figure 1 could influence the occurence of shorter time differences between adjacent speech epochs. To adjust for that possibility, we determined CV2, a more stringent measure of irregularity used to examine regularity of time epochs in brain neuronal activity [11]. A lower value of CV2 indicates higher regularity of adjacent epoch durations. We found that CV2 decreased after a month of improvisational theater training (Figure 4B)

**DISCUSSION**

Our proof-of-concept study offers a possibility that probing speech occurrence patterns can track changes in the evolution of behavior relevant to improvisational theater training. We show that speech epoch durations are stable for a single day but decrease with a month of training. Moreover, the regularity of adjacent speech epoch durations increased with training, suggesting an improvement in sharing focus and speaking time between players in a scene.

*Rhythm and regularity in speech epoch durations*
An a priori expectation was that with training and better understanding between players, there would be an intuitive metronome of speech epoch exchanges depicting a decrease in talking over each other. Contrary to prediction, we found no increase in the rhythm of speech exchanges with training. The possibility that such rhythimic exchanges did occur, but were too few to be reflected in the power spectral density (which averages 20 second segments across an approximate 20 minute period) cannot be ruled out. In support, measuring discrete speech epoch volleys showed an increase in slow frequency speech volley durations with training. Interestingly, another more granular measure of sharing focus, that of matching one's speech duration to the preceding speech duration, did show an improvement with training. Similar to plasticity of speech epoch durations, this measure of regularity was stable across a single day.

*Putative mechanisms*
While speculative, improvisational theater performances arguably involve an adaptive task. An evolving scene requires continual attention and an adaptation to the immedate required response. Adaptive task difficulty modulates neural plasticity and generalization of training [12] and could play a role in the mechanisms involved in behavioral plasticity induced by improvisational theater training. Since improvisational performance often involves empathizing and unconditional support of partners, plasticity in deeper brain structures involved in empathy training could have a contributing role [13].

*Limitations*
One limiting factor is the paucity of observations, thus limiting this study to a proof-of-concept assay of methodology rather than a broadly applicable finding. Another limitation is restricting analysis to the auditory domain. Communication between humans can involve visual and tactile and non-speech sound initiations which are missed in our current study focused on auditory information.

## ACKNOWLEDGEMENTS

We are grateful for the training and collaborative environment provided by Dallas Comedy House, Four Day Weekend, and Stomping Grounds and our multiple coaches over the years. Specifically we are thankful to Todd Upchurch and Gabriel Vasquez who coached troupe Hippo Campus. We thank Molly Jakkamsetti for providing constructive input for the manuscript.